\documentclass[superscriptaddress,preprintnumbers,showpacs,prb,aps,twocolumn]{revtex4-2}

\usepackage{amsfonts}
\usepackage{amssymb}
\usepackage{amsmath}
\usepackage{graphicx}
\usepackage{natbib}
\usepackage{physics}
\usepackage{siunitx}
\usepackage{float}
\usepackage{multirow}
\usepackage{ulem}

\usepackage{fancyhdr}

\setcitestyle{numbers,square}

\usepackage[usenames,dvipsnames]{color}
\usepackage{soul}

\usepackage[hidelinks]{hyperref} 

\usepackage{tablefootnote}
\makeatletter
\newcommand\footnoteref[1]{\protected@xdef\@thefnmark{\ref{#1}}\@footnotemark}
\makeatother

\begin{document}

\newcommand{\ie}{{\it i.e.}}
\newcommand{\eg}{{\it e.g.}}
\newcommand{\etal}{{\it et al.}}

\newcommand{\micron}{$\mu$m}

\newcommand{\Kxx}{$\kappa_{\rm {xx}}$}
\newcommand{\Kxy}{$\kappa_{\rm {xy}}$}
\newcommand{\Kzy}{$\kappa_{\rm {zy}}$}
\newcommand{\Kzz}{$\kappa_{\rm {zz}}$}

\newcommand{\ncco}{Nd$_{2-x}$Ce$_x$CuO$_4$}
\newcommand{\pcco}{Pr$_{2-x}$Ce$_x$CuO$_4$}
\newcommand{\ndlsco}{La$_{1.6-x}$Nd$_{0.4}$Sr$_x$CuO$_4$}
\newcommand{\eulsco}{La$_{2-y-x}$Eu$_{y}$Sr$_x$CuO$_4$}
\newcommand{\ybco}{YBa$_{2}$Cu$_3$O$_y$}

\newcommand{\TN}{$T_{\rm {N}}$}
\newcommand{\Tc}{$T_{\rm {c}}$}

\title{Planar thermal Hall effect from phonons in cuprates}

\author{Lu~Chen}
\email{lu.chen@usherbrooke.ca}
\affiliation{Institut quantique, D\'epartement de physique \& RQMP, Universit\'e de Sherbrooke, Sherbrooke, Qu\'ebec, Canada}

\author{L\'ena Le Roux}
\affiliation{Institut quantique, D\'epartement de physique \& RQMP, Universit\'e de Sherbrooke, Sherbrooke, Qu\'ebec, Canada}

\author{Ga\"el~Grissonnanche}
\affiliation{Institut quantique, D\'epartement de physique \& RQMP, Universit\'e de Sherbrooke, Sherbrooke, Qu\'ebec, Canada}

\author{Marie-Eve~Boulanger}
\affiliation{Institut quantique, D\'epartement de physique \& RQMP, Universit\'e de Sherbrooke, Sherbrooke, Qu\'ebec, Canada}

\author{Steven~Th\'eriault}
\affiliation{Institut quantique, D\'epartement de physique \& RQMP, Universit\'e de Sherbrooke, Sherbrooke, Qu\'ebec, Canada}

\author{Ruixing Liang}
\affiliation{Department of Physics \& Astronomy, University of British Columbia, Vancouver, British Columbia, Canada}
\affiliation{Canadian Institute for Advanced Research, Toronto, Ontario, Canada}

\author{D. A. Bonn}
\affiliation{Department of Physics \& Astronomy, University of British Columbia, Vancouver, British Columbia, Canada}
\affiliation{Canadian Institute for Advanced Research, Toronto, Ontario, Canada}

\author{W. N. Hardy}
\affiliation{Department of Physics \& Astronomy, University of British Columbia, Vancouver, British Columbia, Canada}
\affiliation{Canadian Institute for Advanced Research, Toronto, Ontario, Canada}

\author{S. Pyon}
\affiliation{Department of Advanced Materials Science, University of Tokyo, Kashiwa, Japan}
\affiliation{Department of Applied Physics, University of Tokyo, Tokyo, Japan}

\author{T. Takayama}
\affiliation{Department of Advanced Materials Science, University of Tokyo, Kashiwa, Japan}
\affiliation{Max Planck Institute for Solid State Research, Stuttgart, Germany}

\author{H. Takagi}
\affiliation{Department of Advanced Materials Science, University of Tokyo, Kashiwa, Japan}
\affiliation{Max Planck Institute for Solid State Research, Stuttgart, Germany}
\affiliation{Department of Physics, University of Tokyo, Tokyo, Japan}
\affiliation{Institute for Functional Matter and Quantum Technologies, University of Stuttgart, Stuttgart, Germany}

\author{Kejun~Xu}
\affiliation{Geballe Laboratory for Advanced Materials, Stanford University, Stanford, California, USA}
\affiliation{Stanford Institute for Materials and Energy Sciences, SLAC National Accelerator Laboratory, Menlo Park, California, 94025, USA}
\affiliation{Departments of Physics and Applied Physics, Stanford University, Stanford, California, USA}

\author{Zhi-Xun~Shen}
\affiliation{Geballe Laboratory for Advanced Materials, Stanford University, Stanford, California, USA}
\affiliation{Stanford Institute for Materials and Energy Sciences, SLAC National Accelerator Laboratory, Menlo Park, California, 94025, USA}
\affiliation{Departments of Physics and Applied Physics, Stanford University, Stanford, California, USA}

\author{Louis~Taillefer}
\email{louis.taillefer@usherbrooke.ca}
\affiliation{Institut quantique, D\'epartement de physique \& RQMP, Universit\'e de Sherbrooke, Sherbrooke, Qu\'ebec, Canada}
\affiliation{Canadian Institute for Advanced Research, Toronto, Ontario, Canada}

\date{\today}

\begin{abstract}
A surprising ``planar" thermal Hall effect,
whereby the field is parallel to the current, 
has recently been observed in a few magnetic insulators, 
and this has been attributed to exotic excitations such as Majorana fermions or chiral magnons.
%
%
%
Here we investigate the possibility of a planar thermal Hall effect in three different cuprate materials,
%
%
in which the conventional thermal Hall conductivity \Kxy~(with an out-of-plane field perpendicular to the current) is dominated by either electrons or phonons.
%
%
%
%
%
%
%
%
%
Our measurements show that the planar \Kxy~from electrons in cuprates is zero,
as expected from the absence of a Lorentz force in the planar configuration.
By contrast, 
we observe a sizable planar \Kxy~in those samples where the thermal Hall response is due to phonons,
even though it should in principle be forbidden by the high crystal symmetry.
Our findings call for a careful re-examination of the mechanisms responsible for the phonon thermal Hall effect in insulators.
%

\end{abstract}

\pacs{Valid PACS appear here}

\maketitle

\section{INTRODUCTION}

\begin{figure*}[t]
\centering
\includegraphics[width = 0.9 \linewidth]{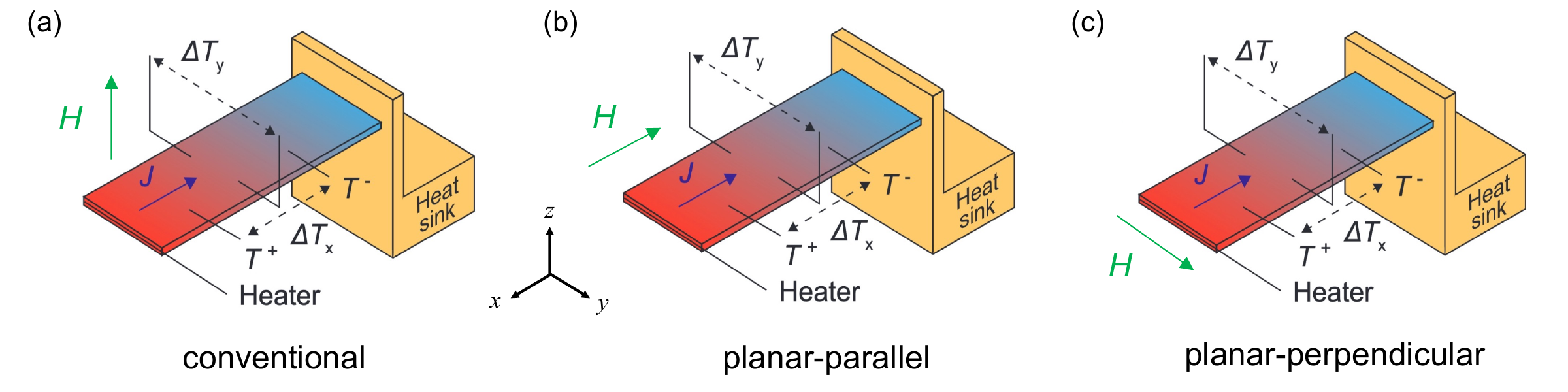}
\caption{
Schematic of the thermal transport measurement setup. (a) Conventional configuration: $H$ $\|$ $z$. (b) Planar-parallel configuration: $H$ $\|$ $x$. (c) Planar-perpendicular configuration: $H$ $\|$ $y$ (see Methods). Directions of both thermal current $J$ and external magnetic field $H$ are shown with colored arrows. Panel (a) is the configuration for measuring the conventional \Kxy. Panel (b) and (c) are for planar \Kxy.
}
\label{fig1}
\end{figure*}

The conventional thermal Hall effect describes the appearance of a transverse temperature gradient along the $y$ direction
when a heat current $J$ is applied along the $x$ direction and a magnetic field $H$ is applied along the $z$ direction.
The thermal Hall effect has become a powerful technique to probe neutral excitations in various quantum materials.
It has been shown both theoretically and experimentally that neutral excitations, 
such as magnons \cite{Onose2010,Katsura2010}, 
phonons \cite{Grissonnanche2019,Grissonnanche2020,Boulanger2020,Li2020,boulanger2022,Chen2022,Uehara2022,Li2023,Ataei2023}, 
or even more exotic quasiparticles like Majorana fermions \cite{Kasahara2018,Bruin2022},
are able to generate a conventional thermal Hall effect.

Recently, the thermal Hall effect measured in a ``planar'' configuration, 
i.e. by applying the field inside the $xy$ plane of the sample (as shown in Fig. \ref{fig1}(b) and (c)), 
has been studied in two Kitaev magnets.
The first planar thermal Hall effect was observed in the Kitaev candidate material $\alpha$-RuCl$_{3}$, 
where there is still a debate on the existence of a half-quantized thermal Hall conductivity \Kxy~and whether the \Kxy~signal comes from Majorana fermions \cite{Yokoi2021} or chiral magnons \cite{Czajka2023,Chern2021,Zhang2021}. 
A planar thermal Hall effect was also observed in the Kitaev candidate material Na$_{2}$Co$_{2}$TeO$_{6}$ \cite{Takeda2022},
where the signal was attributed to chiral magnons.

It has been argued that phonons may in fact be responsible for generating the conventional \Kxy~measured in those two materials \cite{Lefrancois2022,Gillig2023,Yang2022}.
The question then is whether phonons can also produce a planar \Kxy.
Here we address this question by studying the planar \Kxy~in another family of materials, 
the cuprates,
where the conventional thermal Hall effect has been extensively studied \cite{Grissonnanche2016,Grissonnanche2019,Grissonnanche2020,Boulanger2020,boulanger2022}.
We report data on the planar \Kxy~in three different cuprates where the conventional \Kxy~is dominated by either electrons or phonons.
The first is \ybco (YBCO), a superconductor where the conventional \Kxy~is dominated by electrons \cite{Grissonnanche2016,Zhang2001}.
We find that the planar \Kxy~is zero,
as expected due to the absence of Lorentz force in a planar-parallel configuration where $H$ $\|$ $J$ (Fig. \ref{fig1} (b)).

The second material is \ncco~(NCCO), 
an insulator at low doping where the conventional \Kxy~is entirely caused by phonons \cite{boulanger2022}.
Surprisingly, we find a non-zero planar \Kxy~with a configuration of  $H$ $\|$ $J$ $\|$ $a$,
of comparable magnitude to the conventional \Kxy.
%
%
%
%
%

The third cuprate is the superconductor \eulsco~(Eu-LSCO) at dopings $p$ = 0.21 and 0.24, 
that locate right on the two sides of the critical doping $p^{\star} = 0.23$.
It has been shown that phononic conventional \Kxy~only exists inside the pseudogap phase, $i.e.$ when $p<p^{\star}$ \cite{Grissonnanche2020}.
We find that the planar \Kxy~is zero at $p$ = 0.24, 
where the conventional \Kxy~only comes from electrons.
However, the planar \Kxy~is non-zero at $p$ = 0.21, 
where the conventional \Kxy~comes from both electrons and phonons. 
The comparison between Eu-LSCO $p$ = 0.21 and 0.24 indicates that the non-zero planar \Kxy~in $p$ = 0.21 is contributed by phonons.

All three cuprates have high crystal symmetries, 
with YBCO being orthorhombic, NCCO and Eu-LSCO being tetragonal within the temperature range of the measurement,
$i.e.$ below 110 K. 
In theory, a planar \Kxy~is not allowed in any of the systems measured here due to their high crystal symmetry \cite{Kurumaji2023}.
Nevertheless, we observe a sizable planar \Kxy~coming from phonons,
which clearly violates the symmetry requirements. 
Our results impose a re-examination of the mechanisms responsible for the phonon thermal Hall effect in insulators.

\section{METHODS}

\subsection{Samples}

\setlength{\tabcolsep}{4pt}
\begin{table}[!]
  \centering
  \label{Table:thermal_conductivity}
\begin{tabular*}{\linewidth}[t]{ccccc}
    \hline
    \hline
    \noalign{\vskip 0.1cm} 
   Material & doping & $L$ ($\mu$m) & $w$  ($\mu$m)   & $t$  ($\mu$m)   \\
    \hline
    \noalign{\vskip 0.1cm}
 \ybco  & $p=0.11$ & 478 & 784 & 37 \\
 
 \ncco  &$x=0.04$ & 1209 & 1232 &  149 \\

\eulsco  & $p=0.21$ & 652 & 553 & 127  \\

    \eulsco  & $p=0.24$ & 534 & 476 & 168 \\
    
    \noalign{\vskip 0.1cm}
    \hline
    \hline

\end{tabular*}
\caption{Sample information including the doping level and dimensions of the contacts (length between contacts $L$ $\times $ width $w$ $\times$ thickness $t$, in $\mu$m).}
\end{table}

Our detwinned single crystal of \ybco~with $y = 6.54$ was grown at the University of British Columbia by a flux method, as described in ref. \cite{Liang2006}. 
The hole doping level $p = 0.11$ is obtained from a superconducting transition temperature $T_{c}$ = 60.5 K. 
Our single crystal of \ncco~with $x = 0.04$ was grown at Stanford University by the traveling-solvent floating-zone method in O$_2$. 
Single crystals of \eulsco~were grown at the University of Tokyo by the travelling-float-zone techinque,
with a Eu concentration of $y = 0.2$ and nominal Sr concentration of $x = 0.21$ and 0.24. 
The hole concentration is given by $p = x$, with an error bar of $\pm$ 0.005.
The superconducting transition temperature $T_{c}$, 
defined by the zero resistance temperature, 
is 14 K and 9 K for Eu-LSCO $x=0.21$ and 0.24. 
The critical doping level where the pseudogap phase ends in Eu-LSCO is at $p^{\star} = 0.23$ \cite{Michon2019}.
A field of 15 T is sufficient to entirely suppress superconductivity in both samples \cite{Michon2019}.

For the thermal transport measurements, 
crystals were cut and polished into rectangular platelets with the longest edge along the crystal $a$-axis.
For YBCO $p=0.11$ sample, gold contacts are diffused on it.
For NCCO $x=0.04$, Eu-LSCO $x=0.21$ and 0.24, contacts were made using silver epoxy, diffused at 500\si{\celsius} under constant oxygen flow for 1~\si{\hour}.
The dimensions (length between contacts $L$ $\times $ width $w$ $\times$ thickness $t$, in $\mu$m) of all the measured samples are listed in TABLE I.

\subsection{Thermal transport measurements}

The thermal conductivity \Kxx~and thermal Hall conductivity \Kxy~are measured by applying a heat current $J$ along the $x$ axis of the sample (longest direction) 
and a magnetic field $H$ either perpendicular to $J$ ($H$ $\|$ $z$ in Fig. \ref{fig1}(a); $H$ $\|$ $y$ in Fig. \ref{fig1}(c)) or parallel to $J$ ($H$ $\|$ $x$; Fig. \ref{fig1}(b)).
$H$ $\|$ $z$ measures the so-called ``conventional” thermal Hall effect, 
while  $H$ $\|$ $y$ and $H$ $\|$ $x$ measures the so-called ``planar" thermal Hall effect.
The heat current $J$ generates a longitudinal temperature difference $\Delta T_{\rm{x}} = T^{+} - T^{-}$ along $x$ direction. 
The thermal conductivity \Kxx~is defined by
\begin{align}
  \kappa_{\rm{xx}} = \frac{J}{\Delta T_{\rm x}}\left(\frac{L}{wt}\right),
\end{align}
where $w$ is the sample width, $t$ its thickness and $L$ the distance between $T^{+}$ and $T^{-}$.
When a magnetic field is applied along either $x$, $y$ or $z$ direction,
a transverse temperature difference $\Delta T_y$ can be measured along the $y$ axis. 
The thermal Hall conductivity \Kxy~is then given by 
\begin{align}
  \kappa_{\rm{xy}} = -\kappa_{\rm{yy}}\left(\frac{\Delta T_{y}}{\Delta T_{x}}\right)\left(\frac{L}{w}\right).
\end{align}
In a tetragonal system, 
such as NCCO and Eu-LSCO,
we can take $\kappa_{\rm{yy}} = \kappa_{\rm{xx}}$. 
In an orthorhombic system like YBCO, 
$\kappa_{\rm{yy}}$ is measured in a separate sample with the heat current along the $y$ direction \cite{Grissonnanche2016}.

Both thermal conductivity \Kxx~and thermal Hall conductivity \Kxy~are measured with a steady-state method. 
The thermal gradient along the sample is provided by a resistive heater connected to one end of the sample. 
The other end of the sample is glued to a heat sink with silver paint.
The samples are carefully mounted to the heat sink with a misalignment of the in-plane magnetic field of at most $1^{\circ}$.
The longitudinal and transverse temperature differences $\Delta T_x$ and $\Delta T_y$ are measured using type-E thermocouples.
All the measurements are carried out in a standard variable temperature insert (VTI) system up to 15 T.

The measurement procedure is the following:
a magnetic field of $+H$ is applied at $T$ = 90 K,
then the sample is cooled down with the $+H$ field to the base temperature ($\sim$ 2-3 K). 
The $+H$ data was taken by changing temperature in discrete steps at a fixed magnetic field during the warm up process.
After the $+H$ data were taken, 
a magnetic field of $-H$ is applied at the same temperature of $T$ = 90 K,
then the sample is cooled with $-H$ field down to base temperature. 
The $-H$ data were taken by the same method during the warm up process.
During the measurement, 
after the temperature is stabilized at each step, the background of the thermocouple is eliminated by subtracting the heater-off value from the heater-on value. 
The contamination from \Kxx~in \Kxy~due to contact misalignment is removed by doing field anti-symmetrization of the transverse temperature difference $\Delta T_y$. 
That is to say, $\Delta T_y$ are measured with both positive and negative magnetic fields in the same conditions, then \Kxy~is calculated using the field anti-symmetrizaed $\Delta T_y$, $i.e.$ $\Delta T_y (H) = \left[\Delta T_y (T,H) - \Delta T_y (T,-H)\right]/2$.
More details regarding the technique can be found in refs.~\cite{Boulanger2020,Grissonnanche2016,Grissonnanche2020}.

\section{RESULTS}

\begin{figure}[!]
\centering
\includegraphics[width = 0.7 \linewidth]{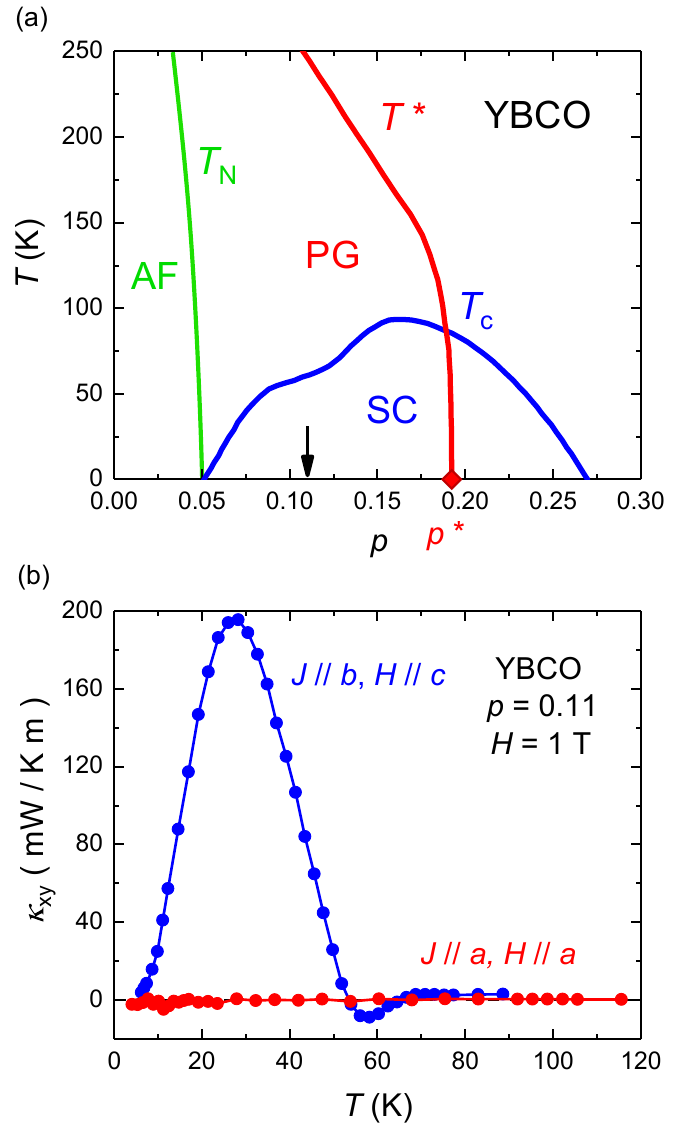}
\caption{
(a) Schematic temperature-doping phase diagram of the hole-doped cuprate YBCO. 
The antiferromagnetic (AF) phase is delineated by the N\'eel temperature $T_{N}$ (green line). 
The superconducting phase (SC) is bounded by the zero-field critical temperature $T_{c}$ (blue line). 
The pseudogap (PG) phase is marked by the critical temperature $T^{\star}$ (red line) and critical doping $p^{\star}$ = 0.19 (red diamond). 
The black arrow marks the doping level of the sample being measured in this study $p$ = 0.11.
(b) Thermal Hall conductivity for YBCO $p$ = 0.11 measured at $H$ = 1 T, 
plotted as \Kxy~vs $T$.
\Kxy~measured with the conventional configuration of $J$ $\|$ $b$, $H$ $\|$ $c$ is marked in blue.
\Kxy~measured with the planar configuration of $H$ $\|$ $J$ $\|$ $a$ is marked in red.
Compared to the large conventional \Kxy~signal, the planar \Kxy~is zero at the same field.
Lines are guide to the eye.
}
\label{fig2}
\end{figure}

\begin{figure}[!]
\centering
\includegraphics[width = 0.75 \linewidth]{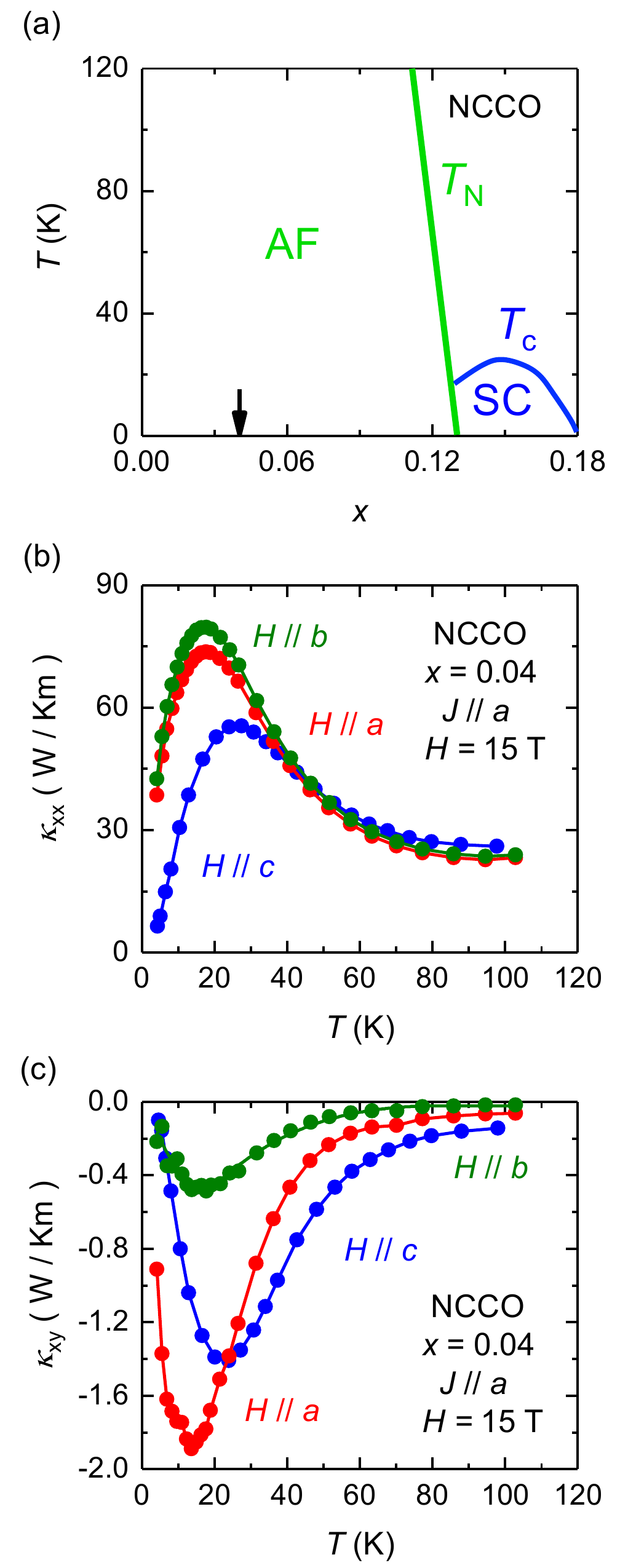}
\caption{
(a) Schematic temperature-doping phase diagram of the electron-doped cuprate NCCO. 
The antiferromagnetic (AF) phase is delineated by the N\'eel temperature $T_{\rm{N}}$ (green line). 
The superconducting phase (SC) is bounded by the zero-field critical temperature $T_c$ (blue line). 
The black arrow marks the doping level of the sample being measured in this study $x$ = 0.04.
(b) Thermal conductivity ((c) thermal Hall conductivity) for NCCO $x$ = 0.04 measured at $H$ = 15 T,
plotted as \Kxx~vs $T$ (\Kxy~vs $T$).
\Kxx~and \Kxy~measured with the conventional configuration of $J$ $\|$ $a$, $H$ $\|$ $c$ (blue) is taken from ref. \cite{boulanger2022}.
\Kxx~and \Kxy~measured with the planar configuration of $H$ $\|$ $J$ $\|$ $a$ ($J$ $\|$ $a$, $H$ $\|$ $b$) is marked in red (green).
The magnitude of the planar-parallel \Kxy~($H$ $\|$ $J$ $\|$ $a$) is comparable to the conventional \Kxy.
%
%
Lines are a guide to the eye.
}
\label{fig3}
\end{figure}

\subsection{YBCO $p$ = 0.11}

First of all, we measured the planar \Kxy~in YBCO $p=0.11$, 
which is a cuprate superconductor with $T_{c} = 60.5$ K (as shown in Fig. \ref{fig2}(a)).
The conventional thermal conductivity \Kxx~and thermal Hall conductivity \Kxy~measured in one sample with a configuration of $J$ $\|$ $b$, $H$ $\|$ $c$ 
at $H$ =  1 T is plotted in Fig. \ref{fig2}(b).
%
%
%
%
%
%
%
%
%
The huge increase in \Kxy~below $T_{c}$ comes from an enhancement of the electronic mean free path due to the loss of inelastic scattering as electrons condense into Cooper pairs, 
in clean crystals where the elastic scattering due to disorder is very small \cite{Zhang2001}. 
The \Kxy~signal in this conventional configuration ($H$ $\|$ $c$) is completely dominated by electrons (more specifically $d$-wave quasiparticles) and these respond to the Lorentz force (given that $H$ $\bot$ $J$). 

In a second sample cut from the same mother crystal, 
we performed the same measurement of \Kxy, 
but with the field applied parallel to the current ($H$ $\|$ $J$ $\|$ $a$). 
The data measured also at $H$ = 1 T is shown in Fig. \ref{fig2} (b). 
We see that \Kxy~$=0$, 
as expected since there is no Lorentz force acting on the electrons (given that $H$ $\|$ $J$). 
This shows that electrons indeed do not generate a planar-parallel Hall effect, 
at least in this orthorhombic cuprate. 
This measurement also demonstrates that our method for detecting a planar \Kxy~is reliable to a high accuracy.

%
%
%

\subsection{NCCO $x$ = 0.04}

Then we turn to the antiferromagnetic (AF) insulating cuprates NCCO at a doping $x$ = 0.04 (Fig. \ref{fig3}(a)), 
where a large conventional \Kxy~contributed by phonons was reported before \cite{boulanger2022}.
The conventional \Kxy~in NCCO $x$ = 0.04 is attributed to phonons for the following reasons.
First, electronic conduction is negligible in this nearly insulating sample.
Secondly, \Kxy($T$) displays a temperature dependence that closely mimics that of the phonon-dominated \Kxx($T$).
Third, when applying the heat current along the $c$ axis (and magnetic field along the $a$-axis), 
the thermal Hall angle \Kzy/\Kzz~shows a comparable magnitude to \Kxy/\Kxx, 
$i.e.$ in the order of $\sim$ 1\%.
As we know that when the heat current is applied perpendicular to the CuO$_{2}$ plane,
the only heat carriers that can propagate along $c$ axis are phonons since magnons are confined inside the CuO$_{2}$ planes.
So \Kzy~is purely contributed by phonons.
The similar thermal Hall angle between \Kxy~and \Kzy~ indicates that \Kxy~is also contributed by phonons.

The temperature-dependent thermal conductivity \Kxx~measured at $H$ = 15 T with three different configurations is plotted in Fig. \ref{fig3}(b)).
\Kxx~shows a strong anisotropy with the field directions,
as has been observed in the mother compound Nd$_{2}$CuO$_{4}$ \cite{Li2005}.
As shown in Fig. \ref{fig3}(c)), 
the conventional \Kxy~taken with $J$ $\|$ $a$, $H$ $\|$ $c$ at $H$ = 15 T (blue curve) shows a temperature dependent negative signal,
which is a typical behavior of phononic \Kxy~observed in both hole-doped and electron-doped cuprates \cite{Grissonnanche2019,Grissonnanche2020,boulanger2022}.
The planar \Kxy~measured in the same sample with a configuration of $H$ $\|$ $J$ $\|$ $a$ and $J$ $\|$ $a$, $H$ $\|$ $b$ are shown in Fig. \ref{fig3}(b)).
The planar-parallel \Kxy~taken with $H$ $\|$ $J$ $\|$ $a$ at $H$ = 15 T (red curve) is non-zero and shows a comparable magnitude to the conventional \Kxy.
%
%
%
For a field in the plane but perpendicular to the current ($H$ $\|$ $b$),
a non-zero planar-perpendicular \Kxy~signal is also observed (green curve in Fig. \ref{fig3}(b)).
The planar \Kxy~measured in both configurations not only follows the same temperature dependence as the \Kxx~signal taken simultaneously,
but also peaks at the same temperature as \Kxx.
This concomitant behavior between \Kxy~and \Kxx~confirms a phononic origin for the planar \Kxy~in NCCO $x$ = 0.04.

\subsection{Eu-LSCO $p$ = 0.21 and 0.24}

\begin{figure}[!]
\centering
\includegraphics[width = 0.7 \linewidth]{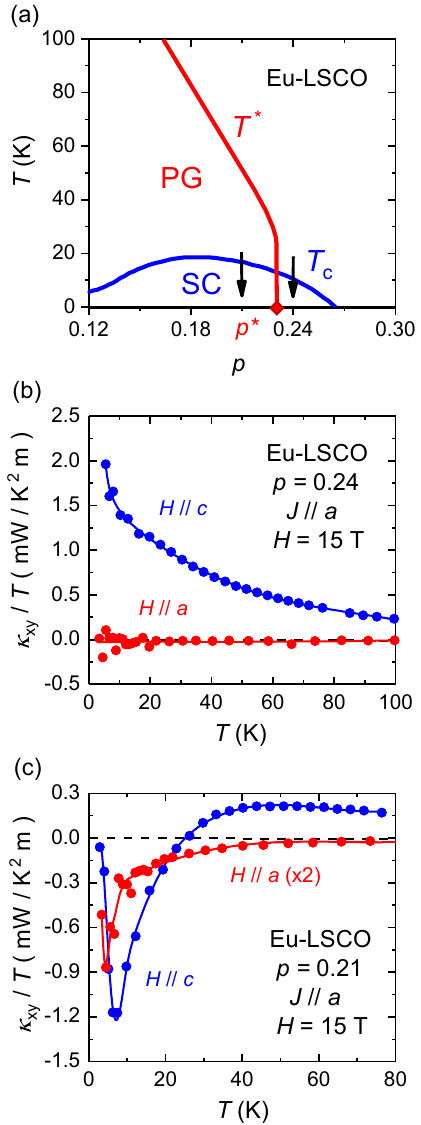}
\caption{
(a) Schematic temperature-doping phase diagram of the hole-doped cuprate Eu-LSCO. 
The superconducting phase (SC) is bounded by the zero-field critical temperature $T_c$ (blue line). 
The pseudogap phase (PG) is bounded by the critical temperature $T^{\star}$, 
which ends at the critical doping $p^{\star} = 0.23$ (red diamond).
The black arrows mark the doping levels of the two Eu-LSCO samples measured in this study, at $p$ = 0.21 and 0.24.
Thermal Hall conductivity for Eu-LSCO at (b) $p$ = 0.24 and (c) $p$ = 0.21, measured at $H$ = 15 T,
plotted as \Kxy/$T$~vs $T$.
\Kxy~measured with the conventional configuration of $J$ $\|$ $a$, $H$ $\|$ $c$ (blue) is taken from ref. \cite{Grissonnanche2019}.
The conventional \Kxy~contains both a positive electron contribution and a negative phonon contribution at $p=0.21$,
while it is purely contributed by electrons at $p=0.24$ outside the pseudogap phase \cite{Grissonnanche2020}.
\Kxy~measured with the planar-parallel configuration of $H$ $\|$ $J$ $\|$ $a$ is marked in red.
The planar \Kxy~is zero at $p$ = 0.24,
where only electrons generate a Hall response,
whereas it is non-zero at $p$ = 0.21,
when phonons contribute.
Lines are a guide to the eye.
}
\label{fig4}
\end{figure}

After studying the planar \Kxy~in materials where the conventional \Kxy~is either dominated by electrons (YBCO $p$ = 0.11) or phonons (NCCO $x$ = 0.04),
we turn to the third cuprate material, Eu-LSCO,
where the phononic conventional \Kxy~can be switched on and off by changing the doping level.
It has been shown in a previous study \cite{Grissonnanche2019,Grissonnanche2020} that
the conventional \Kxy~in the hole-doped cuprates Nd-LSCO and Eu-LSCO are contributed by both electrons and phonons inside the pseudogap phase,
while it is only contributed by electrons when the doping level exceeds the pseudogap critical doping $p^{\star}$.
The phononic origin of the conventional \Kxy~in Eu-LSCO and Nd-LSCO inside the pseudogap phase is established in the following way.
When applying the heat current along the $c$ axis (and magnetic field along the $a$ axis),
\Kzy~is zero at $p$ = 0.24,
while \Kzy~shows a large negative signal at $p$ = 0.21 \cite{Grissonnanche2020}.
Since magnons are highly confined inside the CuO$_{2}$ and the electron contribution to \Kzy~is negligible according to the Wiedemann–Franz Law,
\Kzy~is only contributed by phonons.
These set of experiments indicate that the conventional \Kzy~is purely contributed by phonons and only non-zero inside the pseudogap phase.
When looking back at \Kxy, the non-zero \Kxy~signal at $p$ = 0.24 is contributed only by electrons, 
while \Kxy~at $p$ = 0.21 is dominated by the positive electronic contribution at high temperature and the negative phononic contribution at low temperature (Fig. \ref{fig4} (c)).

The two Eu-LSCO samples measured in this study are $p$ = 0.21 and $p$ = 0.24, 
respectively located on each side of $p^{\star} = 0.23$ (Fig. \ref{fig4}(a)).
In Eu-LSCO $p$ = 0.24, the conventional \Kxy~taken with $J$ $\|$ $a$, $H$ $\|$ $c$ at $H$ = 15 T is plotted as the blue curve in Fig. \ref{fig4}(b),
which is positive and only contributed by electrons.
The planar \Kxy~measured in the same sample with $H$ $\|$ $J$ $\|$ $a$ shows a zero signal (red curve).
This result shows again that in a material
where the conventional \Kxy~is only contributed by electrons,
the planar \Kxy~signal is zero,
in agreement with what we found in YBCO $p$ = 0.11.
In this tetragonal cuprate,
there is no planar-parallel \Kxy~signal.
This second measurement also confirms that our method for detecting a planar \Kxy~is reliable,
this time in a field as large as 15 T. 

In Eu-LSCO $p$ = 0.21, the conventional \Kxy~taken with $J$ $\|$ $a$, $H$ $\|$ $c$ at $H$ = 15 T is plotted as the blue curve in Fig. \ref{fig4}(c).
We can see that it is dominated by the positive electronic \Kxy~at high temperature and the negative phononic \Kxy~at low temperature.
The planar \Kxy~measured in the same sample with $H$ $\|$ $J$ $\|$ $a$ shows a non-zero signal (red curve),
which grows upon cooling in tandem with the emergence of the phononic conventional \Kxy.
We conclude that in this tetragonal cuprate,
a planar-parallel \Kxy~signal does appear and it is associated with phonons.

\section{DISCUSSION}

Our results in YBCO $p$ = 0.11 at $H$ = 1 T and Eu-LSCO $p$ = 0.24 at $H$ = 15 T show that the planar \Kxy~coming from electrons is zero in cuprates,
as expected since electrons feel zero Lorentz force when the magnetic field is parallel to the heat current ($H$ $\|$ $J$).
These two sets of experiments also establish that our measurement procedure is reliable, 
$i.e.$ doing field antisymmetrization on $\Delta T_y$ between $+H$ and $-H$ runs give the expected result.

%
%
Our results in NCCO $x$ = 0.04 and Eu-LSCO $p$ = 0.21 clearly show that phonons are able to generate a sizable planar thermal Hall signal.
%
%
%
%
%
%
%
%
%
The question is how to reconcile our results with the symmetry requirements?
Theories point out that in order to have a non-zero planar \Kxy,
certain symmetry requirements have to be fulfilled to make $H$ $\|$ $J$ different from $H$ $\|$ $-J$ \cite{Yokoi2021,Chern2021,Zhang2021,Kurumaji2023}.
However, we observe a non-zero planar \Kxy~in two tetragonal systems, NCCO and Eu-LSCO (space group I4/mmm),
which clearly violates the symmetry requirements. 

Our speculation is that a non-zero planar \Kxy~signal could potentially result from a locally broken symmetry due to impurities, defects, or domains.
%
%
A possible mechanism for NCCO, an AF insulator with $T_{N}\sim$ 200 K, is the scattering of phonons by antiferromagnetic domains.
This is reminiscent of the nonmagnetic insulator SrTiO$_{3}$,
where the conventional \Kxy~was attributed to the scattering of phonons by antiferrodistortive structural domains \cite{Li2020,Chen2020}.
In cuprates, disorder coming from dopants and oxygen vacancies could also be the source of local symmetry breaking \cite{boulanger2022}.
This type of extrinsic mechanism could potentially work for both conventional and planar configurations of the phonon thermal Hall effect.

Our findings call for a re-evaluation of the previously reported results in the Kitaev candidate materials. 
In $\alpha$-RuCl$_{3}$, a non-zero planar \Kxy~was only observed with a configuration of $H$ $\|$ $J$ $\|$ $a$ but not with $H$ $\|$ $J$ $\|$ $b$ \cite{Czajka2021}
due to the fact that the $C_{2}$ rotational symmetry is preserved only along $b$-axis but not along $a$-axis \cite{Yokoi2021,Czajka2023,Chern2021,Zhang2021}.
This explanation is based on the symmetry properties of the monoclinic structure in the $C2/m$ space group.
However, three different low temperature structures have been reported for $\alpha$-RuCl$_{3}$: monoclinic $C2/m$ \cite{Johnson2015,Cao2016}, 
trigonal $P3_{1}12$ \cite{Plumb2014}, and rhombohedral $R\overline{3}$ \cite{Glamazda2017}.
This controversy could potentially change the symmetry argument for the planar \Kxy~data.
Furthermore, the thermal transport properties have been reported to highly depend on the stacking disorder in the samples, 
which results from different growth techniques and handling processes \cite{Zhang20231, Zhang20232}.
In Na$_{2}$Co$_{2}$TeO$_{6}$ (space group $P6_{3}22$),
a non-zero planar \Kxy~has been observed along both the zigzag and armchair directions \cite{Takeda2022},
which are in principle both forbidden since the crystal structure has $C_{2}$ rotational symmetry along both directions.
The observation of a planar \Kxy~in Na$_{2}$Co$_{2}$TeO$_{6}$ points to an extrinsic mechanism that can locally break the symmetry. 

Note that in addition to cuprates, some of us have also observed a phononic planar thermal Hall signal (comparable in magnitude to the conventional thermal Hall signal)
in both the Kitaev magnet Na$_{2}$Co$_{2}$TeO$_{6}$ \cite{Chen2023} and in the frustrated antiferromagnetic insulator Y-kapellasite \cite{Quentin2023},
thereby further validating the existence of a planar thermal Hall signal coming from phonons.
%
%
%

%
%
%
%
%
%
%
%
%

\section{CONCLUSION}

We report a systematic study of the planar thermal Hall response in three cuprate materials. 
In orthorhombic YBCO (at $p=0.11$), 
the large electronic \Kxy~observed in the conventional configuration ($H$ $\|$ $c$) disappears entirely when the field is applied parallel to the current ($H$ $\|$ $a$), 
as expected in the absence of a Lorentz force when $H$ $\|$ $J$.
In tetragonal Eu-LSCO at $p = 0.24$, 
where the conventional \Kxy~is entirely electronic, 
we also find zero Hall signal in the planar configuration ($H$ $\|$ $J$ $\|$ $a$). 
In dramatic contrast, 
we observe a planar Hall response that is as large as the conventional Hall response in the tetragonal cuprate NCCO (at $x$ = 0.04), 
an insulator in which the Hall response is entirely due to phonons. 
We conclude that phonons can produce a planar thermal Hall effect even in materials where it is forbidden by global symmetry.
%
%
%
We propose that the planar thermal Hall signal emerges from a local symmetry breaking, presumably associated with impurities, defects or domains.
Our findings call for a re-examination of the mechanism responsible for a phonon thermal Hall effect in insulators and also a re-evaluation of the planar thermal Hall conductivity reported in Kitaev candidate materials.

\section{ACKNOWLEDGMENTS}
We thank S.~Fortier for his assistance with the experiments.
L.T. acknowledges support from the Canadian Institute for Advanced Research and funding from
the Institut Quantique,
the Natural Sciences and Engineering Research Council of Canada (Grant No, PIN:123817),
the Fonds de Recherche du Qu\'{e}bec - Nature et Technologies,
the Canada Foundation for Innovation,
and a Canada Research Chair.
This research was undertaken thanks in part to funding from the Canada First Research Excellence Fund.
Z-X.S. acknowledges the support of the U.S. Department of Energy, Office of Science, Office of Basic Energy Sciences, Division of Material Sciences and Engineering, under contract DE-AC02-76SF00515.

\vfill

\bibliography{reference}

\end{document}